\documentstyle[twocolumn,prl,aps,epsfig]{revtex}
\preprint {IMSc-200/02/01}
\begin{document}
\draft
\title{Splitting between quadrupole modes of dilute quantum
gas in a two dimensional anisotropic trap} 
\author{Tarun Kanti Ghosh${}^{1}$ and Subhasis Sinha${}^{2}$}
\address
{1. The Institute of Mathematical Sciences, C. I. T. Campus, Chennai 600 113, India}
\address
{2.  Laboratorie Kastler-Brossel, 24 rue Lhomond, 75005 Paris, France }

\date{\today}
\maketitle

\begin{abstract}

We consider quadrupole excitations of quasi-two dimensional interacting
quantum gas in an anisotropic harmonic oscillator potential at zero
temperature. Using the time-dependent variational approach, we calculate a
few low-lying collective excitation frequencies of a two dimensional
anisotropic Bose gas. Within the energy weighted sum-rule approach, we derive
a general dispersion relation of two quadrupole excitations of a
two dimensional deformed trapped quantum gas. This dispersion relation is
valid for both statistics. We show that the quadrupole excitation frequencies
obtained from both methods are exactly the same.  Using this general
dispersion relation, we also calculate the quadrupole frequencies of a
two dimensional unpolarized Fermi gas in an anisotropic trap. For both cases,
we obtain analytic expressions for the quadrupole frequencies and the
splitting between them for arbitrary value of trap deformation. This
splitting decreases with increasing interaction strength for both statistics.
For a two dimensional anisotropic Fermi gas, the two quadrupole frequencies 
and the splitting between them become independent of the particle number within the
Thomas-Fermi approach. 

\end{abstract}

\pacs{PACS numbers:03.75.Fi, 05.30Jp}

\narrowtext

\section{introduction} In a series of experiments, Bose-Einstein condensates
(BEC) have been produced by cooling a vapor of alkali atoms  to a 
temperature of a few
nanokelvin \cite{anderson}. This system opens up interesting
perspectives in the field of many body physics. There has been much progress
in the theoretical understanding of this system \cite{dal}. In particular, the
low-energy collective excitation spectrum of a Bose condensed dilute gas in a
trap has been discussed analytically by Stringari \cite{strin} using
a hydrodynamic approximation and also by  sum-rule approach. A few
low-lying excitations have also been calculated analytically by using
time-dependent variational approach \cite{gar}.  The low-energy excitation
spectrum obtained by using time-dependent gaussian variational ansatz exactly
coincides with the hydrodynamic results in the limit of large particle number
N. Also a similar type of scaling ansatz has been used to describe the time
evolution of the condensate in the large N limit \cite{castin,kagan,olsha}.
Experimentally the low-lying collective excitation frequencies of a
condensate have been measured both at zero temperature \cite{mew} and at finite
temperature \cite{jin}. These observed values of the collective oscillation
frequencies are in agreement with theoretical results at zero
temperature. 

After the discovery of BEC in alkali atomic gas, the behaviour of trapped
Fermi gas is also in focus. It is also possible to trap the Fermionic atoms
at very low temperature, where the quantum effects can be observed. There has
been
 experimental progress towards cooling a Fermi gas into the degenerate regime
$ (T< T_F )$ \cite{jin1}. Several authors have studied the thermodynamic
properties \cite{butt,salas}, collective excitation frequencies in the
normal phase \cite{amo,vichi,clark} as well as in the superfluid phase
\cite{tosi,baranov} of a three dimensional trapped Fermi gas. 

The reduction in dimension of a quantum system is the subject of extensive
studies in trapped Bose systems \cite{petrov,ma,pit,kanti,piju} as
well as trapped Fermi systems \cite{sank}. With present technology one can
freeze the motion of the trapped particles in one direction to create a quasi 
two dimensional quantum gas by tuning the frequency in the $ z $ direction. Recently,
the lower dimensional BEC has been realized \cite{gorlitz,burger}. 

In a deformed quantum system, angular momentum is not a good quantum number
and the angular momentum states mix with each other.  Also in the presence of
a small deformation of trap, the states with angular momentum quantum number
$+l$ and $-l$ split.  The partition function of a deformed two-dimensional
harmonic oscillator is exactly the same as the partition function of a rotating
harmonic oscillator, where the rotation frequency $\Omega$ is related to the
trap anisotropy, $\Omega = (\omega_{y} - \omega_{x})/2 $. Here $\omega_{x}$
and $\omega_{y}$ are the oscillator frequencies of the deformed oscillator. 
From this analogy we find that the splitting occurs between the $+l$ and
$-l$ angular momentum states and it is proportional to the trap anisotropy.
Similarly for a quantum gas in a deformed trap, the degenerate multipole modes
with angular momentum quantum number $+l$ and $-l$ also split. The dipole
excitation frequencies of a quantum gas in deformed harmonic trap are the
trap frequencies along x and y directions, $\omega_{x}$ and $\omega_{y}$. The
frequencies and the splitting of the dipole mode neither depends on the
statistics of the trapped gas nor on the interaction between the particles. 
In circular symmetric trap two quadrupole frequencies are degenerate and the degeneracy
is lifted up by elliptic trap deformation.
In the present work we consider quadrupole modes of quantum gas in a quasi
two dimensional anisotropic trap. 
We  calculate the frequencies
of quadrupole like modes for both bosons and fermions in a deformed harmonic
trap analytically. For both statistics, we study the effect of interaction 
on the splitting between the quadrupole modes for arbitrary deformation of the trap.
First we consider the case of a two dimensional deformed trapped Bose gas at
zero temperature. We analyze the nature of  small oscillations of the confined 
gas within the time-dependent variational method. In case of deformed trap the
quadrupole mode couples with the monopole mode whereas the scissors mode 
remains decoupled. This is due to fact that the Hamiltonian is invariant under
reflection. We calculate the 
frequencies of the mixed type of modes using variational technique.  Next we
construct the most general type of excitation operator which couples the
quadrupole and monopole mode.  Then we consider only the minimum energy
excitation in sum rule approach. The excitation energy obtained from the
sum-rule approach exactly matches with that obtained from the time-dependent
variational technique even for an arbitrary number of particles. In the case of a
symmetric trap, these modes can be identified as a monopole mode and a
quadrupole mode.
There is a possibility of experimental verification of our results since the
quasi two dimensional Bose condensed has been realized in MIT \cite{gorlitz}.

This sum-rule method allows us to calculate the monopole mode and the quadrupole
modes of a Fermi system in a deformed trap where the equation of motion
technique does not hold. These modes can be observed in nano structures like
quantum dots. When the two-body interaction is long ranged, the sum rule
method can be generalised. 

There has been no systematic theoretical study on the collective excitations
of a two dimensional deformed trapped quantum gas at zero temperature.  The
purpose of this paper is to give an analytic description of the quadrupole
excitation frequencies of a two dimensional deformed trapped quantum gas at
zero temperature and to calculate the splitting of the quadrupole modes for
an arbitrary deformation of the trap. 

The paper is organised as follows.  In Sec.II, we model the quasi-two
dimensional trapped Bose system.  Using the time-dependent variational method
we calculate the monopole and quadrupole excitation frequencies of a two
dimensional deformed trapped interacting Bose gas. In Sec.III, using the
sum-rule approach we derive a general dispersion relation for the quadrupole
excitation frequencies of a two dimensional trapped quantum system
interacting through the two-body potential. This relation is valid for both
Fermi and Bose statistics. We apply this general dispersion relation to 
calculate the
same excitation frequencies of a trapped Bose system. We show that the
quadrupole excitation frequencies obtained from both methods are exactly
the same. In Sec.IV, we consider  trapped unpolarized fermions and apply the
dispersion relation obtained from sum-rule approach to calculate the
frequencies of quadrupole modes. In Sec. IV, we present the  summary and
conclusions of our work. 

\section{ Collective low-energy excitation frequencies of a two dimensional
deformed trapped Bose gas}

In BEC experiments, the trap potential can be approximated by an effective
three dimensional harmonic oscillator potential, with tunable trap
frequencies $ \omega_{z} $ in the axial (z) direction and $\omega_{x}$,
$\omega_{y}$ in the transverse (x-y) plane. The alkali-metal vapors used in
experiment are very dilute and the interparticale interaction is well
described by the short range pseudopotential and the interaction strength is
determined by s-wave scattering length $a$. Here we consider the case when
the interparticle interaction is strongly repulsive. The Gross
-Pitaevskii(GP) \cite{gross} energy functional of the trapped boson of mass $
m $ is given by,
 \begin{eqnarray} 
E[\psi] & = & \int d^{2}r dz [
\frac{\hbar^2}{2 m} | \nabla \psi(\vec r,z) |^2 \\ \nonumber & + & \frac{ m }
{ 2 } ( \omega_x^2 x^2 + \omega_y^2 y^2 + \omega_z^2 z^2 ) |\psi(\vec r,z)
|^2 \\ \nonumber & + & \frac{g}{2} \int d^2 r^{ \prime} dz^{\prime } \delta
^2 (\vec r - \vec r^{ \prime }) \delta (z - z^{\prime })|\psi (\vec r^{\prime
},z^{\prime}) |^2 |\psi (\vec r, z)|^2 ],
 \end{eqnarray} 
where, $ g = \frac{4
\pi a \hbar^2}{m} $, $\vec r $ is the position vector in $x-y$ plane and $
\psi (\vec{r},z) $ is the condensate wave function. 

It has been shown by Baganato {\em et al}. \cite{baganato} that for an ideal
two dimensional Bose gas under harmonic trap, a macroscopic occupation of the
ground state can exist at temperature $ T < T_c = \sqrt {N / \zeta (2)} \hbar
\omega / k_B $. With  present technology it is possible to freeze the motion of
the trapped particles in one direction to create a quasi-two dimensional Bose
gas. In the frozen direction the particles execute  zero point motion. To
achieve this quasi-two dimensional system, the frequency in the frozen
direction should be much larger than the frequency in the x-y plane and the
mean interactions between the particles.
Alternatively, the trap frequencies are such that  $ \hbar \omega_z \gg \mu \geq \hbar 
\omega_0 $ and $ k_B T \ll \hbar \omega_z $, where $ \mu $ is the chemical potential of
the two dimensional Bose gas. 

For a quasi two dimensional system we may assume that the wave function in the
z-direction is seperable and is given by,
 \begin{equation} 
\psi ( z ) =
\frac{1}{(\sqrt{\pi} a_{z})^{1/2}}e^{- \frac{ z^2 }{2 a_{z}^2 }},
\end{equation}
 where $ a_{z} = \sqrt{ \hbar/m \omega_{z}} $ is the oscillator
length in the z-direction. Now we integrate out the z-component in the three
dimensional GP energy functional, then we get the effective energy functional
in two dimension:

\begin{eqnarray} 
E - \frac{ \hbar \omega_{z} N }{ 2 } & = & \int d^2 r [
\frac{\hbar^2}{2 m} | \nabla \psi(\vec r) |^2\\ \nonumber &  + & 
\frac{ m }{ 2 }(\omega_x^2
x^2 + \omega_y^2 y^2) |\psi(\vec r) |^2 \\ \nonumber & + & \frac{g_{2}}{2}
 \int d^2r^{\prime} \delta ^2 ( \vec r - \vec r ^{\prime} )|\psi(\vec
r^{\prime}) |^2 | \psi (\vec r ) |^2], 
\end{eqnarray}
 where $ g_{2} = 2
\sqrt{2 \pi } \hbar \omega_{z} a_{z} a $ is the effective coupling strength
in two dimension, $ a $ is the $ s $-wave scattering length in three
dimension and N is the total number of particles in the condensate.  The same
effective coupling constant is obtained in Ref. \cite{ma}. The effective
interaction in two dimension is given by, 
\begin{equation} 
V_{I} =
g_{2}\delta^{2}(\vec{r} -\vec{r'}). 
\end{equation}
The chemical potential of quasi two dimensional Bose condensed state is
$ \mu = \hbar \omega_0 \sqrt{\sqrt{2 \pi} N a/a_z} $.
Recently, the two dimensional Bose condensed state has been realized in MIT
\cite{gorlitz}. In this experimental set up, they have loaded $ N \sim 10^4 $
number of sodium atoms in a trap with trap frequencies $ \omega_z/2 \pi = 790 $ Hz,
 $ \omega_0/2 \pi = \sqrt{\omega_x \omega_y} \sim 20 $Hz and
 $ a = 2.75 $ nm. 
One can easily calculate the chemical potential $ \mu \sim 0.19 \hbar \omega_0 \sqrt{N} $
which satisfies the above mentioned inequality condition to be a quasi
two dimensional Bose system.  

In two dimensions, the equation of motion of the condensate wave function is
described by the Gross-Pitaevskii equation, 
\begin{equation}\label{non} 
i\hbar
\frac{\partial \psi (\vec r)}{\partial t} = [- \frac{\hbar^{2}}{2m}\nabla^{2}
+ V(\vec{r}) + g_{2}|\psi(\vec r)|^{2}] \psi(\vec r), 
\end{equation} 
where $ V(\vec{r}) = \frac{1}{2} m (\omega_{x}^2 x^{2} + \omega_{y}^2 y^{2})$ is the
deformed trap potential in two dimension. The normalization condition for
$\psi$ is $ \int d^{2}r |\psi|^{2} = N $. N is the number of particles in the
condensate . One can write down the Lagrangian density corresponding to
this system as follows: 
\begin{eqnarray}\label{density} 
{\cal L} & = & \frac{ i
\hbar }{ 2 }(\psi\frac{\partial{\psi^{*}}}{\partial{ t }} -
\psi^{*}\frac{\partial\psi}{\partial t}) \\ \nonumber & + &
(\frac{\hbar^{2}}{2m} |\nabla \psi
|^{2} + V(\vec{r})|\psi |^{2} + \frac{g_{2}}{2} |\psi |^{4}), 
\end{eqnarray}
where * denotes complex conjugation. One can get the non-linear Schroedinger
equation (\ref{non}) by minimizing the action related to the above Lagrangian
density (\ref{density}). In order to obtain the evolution of the condensate
we assume the most general Gaussian wave function, 
\begin{equation}\label{wave}
 \psi(X,Y,t) = C(t) e^{-\frac{1}{2}[ \alpha (t) X^2 + \beta (t) Y^2 + \gamma
(t) X Y]}, 
\end{equation} where C(t) is the normalization constant. X and Y
are the dimensionless variables, $ X = \frac{x}{a_{0}}$, $ Y =
\frac{y}{a_{0}}$ where $ a_{0} = \sqrt{\frac{\hbar}{m\omega_{0}}} $ is the
oscillator length and $ \omega_0 = \sqrt{\omega_x \omega_y} $ is the mean
frequency.  Further, $ \alpha = \alpha_1 + i \alpha_2 $, 
$ \beta = \beta_1 + i \beta_2
$ and $ \gamma = \gamma_1 + i \gamma_2 $ are the time dependent dimensionless
complex variational parameters. The $ \alpha_1 $ and $ \beta_1 $ are 
inverse square of the condensate widths in $ x $ and $ y $ direction
respectively. The square of the normalization constant is $| C(t) |^2 =
\frac{ N \sqrt{D}}{\pi a_{0}^2 } $, where $ D = \alpha_{1} \beta_{1} -
\gamma_{1}^2 $. The Gaussian ansatz (eq. (\ref{wave})) for the order
parameter can also be generalized to three dimensional anisotropic
trapped Bose system to study various scissors modes. 

The gaussian variational ansatz becomes an exact ground state in the non
interacting limit and in the presence of repulsive interaction it gives rise
to spreading of the condensate wave function. To describe the quadrupoles
and monopole oscillation, we consider the most general time dependent quadratic
exponent of the variational ansatz. 
 
We obtain the effective Lagrangian $ L $ by substituting Eq. (\ref{wave})
into Eq. (\ref{density}) and integrating the Lagrangian density over the
space co-ordinates,
\begin{eqnarray} 
\frac{L}{N\hbar\omega_{0}} & = &
\frac{1}{4 D } [-(\beta_{1}\dot{\alpha_{2}}+\alpha_{1}\dot{\beta_{2}} - 2
\gamma_1 \dot{\gamma_2}) \\ \nonumber & + & (\alpha_{1} + \beta_{1} ) D
 +  (\alpha_{2}^{2} + \gamma_{2}^2 ) \beta_{1}
 +  ( \beta_{2}^2 + \gamma_{2}^2 ) \alpha_{1} \\ \nonumber & - & 2 (
\alpha_{2} + \beta_{2} ) \gamma_{1} \gamma_{2}
 +  \lambda \beta_{1} + \frac{\alpha_{1}}{ \lambda } + P D^{3/2}],
\end{eqnarray}
 where $ \lambda $ is the asymmetric ratio, $ \lambda =
\omega_{x}/ \omega_{y} $ and $ P = \sqrt{\frac{2}{\pi}}2N\frac{a}{a_{z}} $. 

The variational energy of the static condensate at equilibrium is given  in
terms of the equilibrium values of the inverse square width of the condensate
along x and y directions,
\begin{equation} 
\frac{ E}{N\hbar \omega_0} = \frac{1}{4} [ (\alpha_{10} +
\beta_{10} ) + ( \frac{\lambda}{\alpha_{10}} + \frac{1}{\lambda \beta_{10}})
+ P \sqrt{\alpha_{10} \beta_{10}}].
 \end{equation}
One can get the equilibrium value of the variational parameters, $
\alpha_{10} $ and $ \beta_{10} $ by minimizing the energy with respect to the
variational parameters,
\begin{equation} 
\alpha_{10}^{2} = \lambda - \frac{P}{2} \alpha_{10} \sqrt{
\alpha_{10}\beta_{10}},
 \end{equation} 
\begin{equation} 
\beta_{10}^{2} =
\frac{1}{ \lambda } - \frac{P}{2} \beta_{10} \sqrt{ \beta_{10} \alpha_{10}}.
\end{equation} 
From the above two relations, we obtain
 
\begin{equation}\label{width}
 \eta^4 + \frac{P \eta^3 }{2} - \frac{P \eta}{ 2
\lambda^2}-\frac{1}{\lambda^2} = 0, 
\end{equation}
 where $ \eta $ is the ratio
of the condensate widths in the $x$ and $y$ direction, $ \eta =
\sqrt{\frac{\beta_{10}}{\alpha_{10}}} $. From Eq. (\ref{width}) one can say
how $ \eta $ changes with the number of atoms N and the coupling constant
$g_{2}$. The variation of $ \eta $ with the the dimensionless effective
interaction strength $ P $ is shown in Fig. 1. The ratio between the widths
of the condensate $\eta$ varies from $1/\sqrt{\lambda}$ to $1/\lambda$, as
the interaction strength increases from zero to large value (Thomas-Fermi
limit).  In the Thomas-Fermi limit, the equilibrium values of the parameters
$\alpha_{1}$ and $\beta_{1}$ are,
\begin{equation} 
\alpha_{10} = \lambda
\sqrt{ \frac{2}{P} }, \beta_{10} = \frac{1}{\lambda} \sqrt{ \frac{2}{P}}
\end{equation}
 In this limit, the energy per particle is $ \frac{ E }{ N} =
\hbar \omega_{0} \sqrt{\frac{P}{2}} $. In the non-interacting limit, $
\alpha_{10}^2 = \lambda $ and $ \beta_{10}^2 = 1/\lambda $. The energy per
particle is $ \frac{ E }{ N} = \frac{ \hbar \omega_0}{2} ( \sqrt{\lambda} +
\sqrt{1/ \lambda} )$.

We are interested in the low-energy excitations of a Bose system. The
low-energy excitations of the condensate correspond to the small
oscillations of the cloud around the equilibrium configuration. Therefore we
expand the time dependent variational parameters around the equilibrium
points in the following way, 
$ \alpha_{1} = \alpha_{10} + \delta{\alpha_{1}} $, $\beta_{1} = \beta_{10} +
\delta{\beta_{1}} $ and $\alpha_{2}= \delta \alpha_2, \beta_{2}= \delta
\beta_2 $ $ \gamma_{1} = \delta{\gamma_{1}} $, $ \gamma_{10} = 0 $ and $
\gamma_2 = \delta \gamma_2 $. 

Using the Euler-Lagrange equation, the time evolution of the inverse square
of the width around the equilibrium points are given by,
\begin{equation}\label{fluc1} 
\delta\ddot{\alpha_{1}} + \lambda \frac{ (8 + 3
P \eta )} {(2 + P \eta )} \delta \alpha_{1}
 + \frac{P \lambda \eta } {( 2 + P \eta )} \delta \beta_{1} = 0,
\end{equation}
\begin{equation}\label{fluc2} 
\delta\ddot{\beta_{1}} + \frac{ P }{ \lambda (2
\eta + P )} \delta \alpha_{1}
 + \frac{ (8 \eta + 3 P )} {\lambda ( P + 2 \eta ) } \delta \beta_{1} = 0,
\end{equation}
\begin{equation}\label{fluc3} 
\delta\ddot{\gamma_1} + [\frac{4 \lambda \eta^2
}{(2 + P \eta )} + (\lambda + \frac{1}{\lambda })] \delta \gamma_1 = 0.
\end{equation} 
From the above equations we can see that the modes
corresponding to the fluctuations of the average of $x^{2}$ and $y^{2}$ are
coupled, but the mode associated with the fluctuation of the average value of
$xy$ is decoupled. The Eqs. (\ref{fluc1}) and ( \ref{fluc2}) are coupled
equations of the modes $ \alpha_1 $ and $ \beta_1 $, where the mode $ xy $
is decoupled. This is due to the fact that the Hamiltonian is invariant under
reflection, $ x \rightarrow -x $ and $ y \rightarrow y $ or $ x \rightarrow x
$ and $ y \rightarrow - y $, and the modes which are odd or even under this
operation  separate out. 

Now we look for  time dependent solutions of $ e^{i\omega t} $ type, we
obtain from Eqs. (\ref{fluc1}) and (\ref{fluc2}),
\begin{eqnarray}\label{mono1}
 \frac{\omega_{\pm}^2 }{\omega_{0}^2} & = &
\frac{ \lambda }{2} \frac{ (8 + 3 P \eta )}{ ( 2
 + P \eta )} + \frac{1}{ 2 \lambda } \frac{ (8 \eta + 3 P)}{( 2 \eta + P)}
 \\ \nonumber & \pm & \sqrt{[\frac{ \lambda ( 8 + 3 P \eta )}{2 ( 2 + P \eta
)} - \frac{(8 \eta + 3 P )}{2 \lambda ( 2 \eta + P )}]^2 + ( \frac{ P \lambda
\eta^2 }{2 + P \eta })^2}. 
\end{eqnarray}

For an isotropic trap, $ \omega_{+} = 2 \omega_0 $ and $ \omega_{-}^2 = [
\omega_{0}^2( 8 + 2P)]/(2+P)$.  For large $N$ limit, $ \omega_{-} = \sqrt{2}
\omega_{0} $. So $ \omega_{+} $ and $ \omega_{-} $ may be identified as the
monopole mode frequency and quadrupole mode frequency respectively. The
monopole mode is coupled with the quadrupole mode in an anisotropic trap.
However, the monopole mode frequency in an isotropic trap is independent of
the interaction strength of the two-body potential and the number of
particles in the condensate state. This is due to the underlying
$ SO(2,1) $ symmetry in the Hamiltonian \cite{pit}, \cite{kanti}. 

From Eq.(\ref{fluc3}) we obtain, 
\begin{equation}\label{qu1}
\frac{\omega_{s}^2}{\omega_{0}^2} = \frac{4 \lambda \eta^2 }{(2 + P \eta )} +
(\lambda + \frac{1}{\lambda }).
 \end{equation} 
In an isotropic trap, $
\omega_{s} $ becomes $ \omega_{-} $.  In the non-interacting limit, $
\omega_s = \omega_x + \omega_y $. In the Thomas-Fermi limit, Eq. (\ref{qu1})
reduces to $ \omega_{s} = \sqrt{ \omega_{x}^{2} + \omega_{y}^{2}} $. In an
isotropic trap this mode corresponds to the quadrupole excitation. This
excitation is also known as scissors mode \cite{guery}, and this oscillation
has been observed experimentally \cite{mara}. 

\section{ Sum-Rules and Collective excitations}

In this section, we study the quadrupole excitations of a two dimensional
deformed trapped quantum gas
 at zero temperature within the sum rule approach. In the collisionless
regime the collective excitation frequencies of a confined gas are well
described by the sum rule method.  The collective excitation of any system is
usually probed by applying external fields. Given an excitation operator $F$,
many useful quantities of the excited system can be calculated from the
so-called strength function \cite{bohi},
 \begin{equation} 
S_{\pm}(E) = \sum_n
|<n| F_{\pm} |0>|^2 \delta(E-E_{n}), 
\end{equation} where $ E_{n} $ and $ | n > $ are the excitation energy and 
the excited state respectively, and $F_{+}
= F$, $F_{-} = F^{\dag}$. Various energy weighted sum rules are derived
from the moments of the strength distribution function, 
\begin{equation}
m^{\pm}_{k} = \frac{1}{2}\int E^{k}(S_{+}(E) \pm S_{-}(E))dE. \end{equation}
It is easy to see that, for a given $ k $, the moments may be expressed in
terms of the commutators of the excitation operator F with the many body
Hamiltonian $H$. We give below some of the useful energy weighted sum rules,
\begin{equation} m^{-}_{0} = \frac{1}{2} <0|[F^{\dag}, F]|0> , \end{equation}
\begin{equation} m^{+}_1 = \frac{1}{2} <0|[F^{\dag} , [H, F]]|0> ,
\end{equation} \begin{equation} m^{-}_{2} = \frac{1}{2} <0|[J^{\dag}, J]|0> ,
\end{equation} \begin{equation} m^{+}_3 = \frac{1}{2} <0|[[F^{\dag} , H],[H,
[H, F]]]|0> , J = [ H, F], \end{equation} where $[,]$ denotes the commutator
between corresponding operators. Near the collective excitation frequency the
strength distribution becomes sharply peaked, and the collective excitation
energy is described by the moments of the strength distribution,
\begin{equation} \hbar \omega = \sqrt{\frac{m^{+}_3}{m^{+}_1}}\label{coll}.
\end{equation}

Following Ref.\cite{lip}, one can derive the above form of collective
excitation energy by using the variational principle. Given the many body
ground state it is possible to find out the collective excitation energy and
the excited state, if one is able to find an operator $O^{\dag}$, which
satisfies the following equation of motion : \begin{equation} [\hat{H},
O^{\dag}] = \hbar \omega_{c} O^{\dag}. \end{equation} The excitation energy
is then given by the following expression, \begin{equation} \hbar \omega_{c}
= \frac{<0|[O,[\hat{H}, O^{\dag}]]|0>}{<0|O, O^{\dag}|0>}. \end{equation} We
may now take the variational ansatz for $O^{\dag}$ as, $O^{\dag} = F + b J$
with the variational parameter $b$. By minimizing the energy with respect to
the variational parameter, we obtain the collective excitation energy as
$E_{c} = \sqrt{m^{+}_{3}/m^{+}_{1}}$, which is same as eqn.(\ref{coll}). 

Similarly, we construct the most general excitation operator $ F = x^2 +
b y^2 $ when monopole and quadrupole modes are coupled. $ b $ is a
variational parameter. In symmetric trap potential, if $ b = 1 $, F is
monopole mode and if $ b = - 1$, F is the quadrupole mode. In the same
way we can calculate the lowest energy excitation in this particular sector
of excitations variationally. The lowest energy mode turns out to be
the quadrupole mode. 

Calculating the moments $ m_{1} $ and $ m_{3} $ by taking the excitation
operator with the Hamiltonian, we obtain, 
\begin{equation}\label{mini}
E_{coll}^2 = \frac{4 \hbar^2}{m} \frac{ E_x }{< x^2> } \frac{(1 + A b^2  +
B b ) }{( 1 + C b^2  )},
 \end{equation} 
where $ A = \frac{ E_y }{ E_x },
B = \frac{ E_{int}}{2 E_x }, C = \frac{<y^2>}{<x^2>} $. and $ E_x = <T_x> +
<V_x> + \frac{1}{4} <E_{int}> $, $ E_y = <T_y> + <V_y> + \frac{1}{4}
<E_{int}> $. Here $ < > $ denotes the expectation value of the corresponding
operators in the ground state and
 $ T_{x} $, $ V_{x} $ and $T_{y}$, $V_{y}$ represents the kinetic energy and
potential energy along $x$ and $y$ coordinates respectively. The interaction
energy is given by $ E_{int} = g_2/2 \int |\psi |^4 d^2r $. Now we minimize
this collective energy with respect to the variational parameter $ \beta $.
The value of $ \beta $ for which the collective energy is minimum, is given
by 
\begin{equation} 
b_0 = \frac{-2(C - A) \pm \sqrt{4(C-A)^2 + 4 B^2 C
}}{2 B C }. 
\end{equation}
 It can be easily shown that for an isotropic trap,
$ b_0 = \pm 1 $. So we have identified that the monopole mode can be
excited by the operator $ F = x^2 + b_0 y^2 $ and quadrupole mode can be
generated by $ F = x^2 - b_0 y^2 $. 
 Inserting $ b_0 $ into Eq. (\ref{mini}), we obtain the following
collective oscillation frequencies: 
\begin{eqnarray}\label{sum1} 
\omega_{\pm}^2
& = & \frac{2}{m}[(\frac{E_y}{<y^2>} + \frac{E_x}{<x^2>}) \\ \nonumber & \pm
& \sqrt{(\frac{E_y}{<y^2>} - \frac{E_x}{<x^2>})^2 + \frac{E_{int}^2}{4
<x^2><y^2>}}].
 \end{eqnarray} 
Using the variational wave function of the
ground state in deformed trap, one can easily get the excitation frequencies.
The lowest energy excitation frequency in this sector is
\begin{eqnarray}\label{mono2} 
\frac{\omega_{-}^2 }{\omega_{0}^2} & = & \frac{
\lambda }{2} \frac{ (8 + 3 P \eta )}{ ( 2
 + P \eta )} + \frac{1}{ 2 \lambda } \frac{ (8 \eta + 3 P)}{( 2 \eta + P)}
 \\ \nonumber & - & \sqrt{[\frac{ \lambda ( 8 + 3 P \eta )}{2 ( 2 + P \eta )}
- \frac{(8 \eta + 3 P )}{2 \lambda ( 2 \eta + P )}]^2 + ( \frac{ P \lambda
\eta^2}{2 + P \eta })^2}.
\end{eqnarray} 
It can be identified as quadrupole
mode since in an isotropic trap, its excitation frequency exactly matches
with the quadrupole mode frequency. The above expression for the excitation
frequency Eq. (\ref{mono2}) is exactly same as the mode frequency $
\omega_{-} $ in Eq. (\ref{mono1}). 

The higher energy excitation exactly matches within the monopole mode,
although it is not the local minimum of the energy, Eq. (\ref{mini}). The
dispersion relation for this monopole mode frequency is:
\begin{eqnarray} 
\frac{\omega_{+}^2 }{\omega_{0}^2} & = & \frac{
\lambda }{2} \frac{ (8 + 3 P \eta )}{ ( 2
 + P \eta )} + \frac{1}{ 2 \lambda } \frac{ (8 \eta + 3 P)}{( 2 \eta + P)}
\\ \nonumber & + & \sqrt{[\frac{ \lambda ( 8 + 3 P \eta )}{2 ( 2 + P \eta )}
- \frac{(8 \eta + 3 P )}{2 \lambda ( 2 \eta + P )}]^2 + ( \frac{ P \lambda
\eta^2}{2 + P \eta })^2}.
 \end{eqnarray}
 
For another quadrupole mode, the excitation operator is $ F = xy $. Using the
commutation relation, we obtain, 
\begin{equation} m_{1} =
\frac{\hbar^{2}}{2m}<(x^{2} + y^{2})>,
 \end{equation} 
\begin{equation} m_{3} =
m_{3}(T) + m_{3}(v) + m_{3}(ee) ,
\end{equation} 
where, 
\begin{equation}
m_{3}(T) = \frac{\hbar^{4}}{m^{3}}<(p_{x}^{2} + p_{y}^{2})>, 
\end{equation}
\begin{equation}
 m_{3}(V) = \frac{\hbar^{4}}{2m}(\omega_{x}^{2} +
\omega_{y}^{2})<(x^{2} + y^{2})>, 
\end{equation}
\begin{eqnarray} 
m_{3}(ee) & = & \frac{ g_2 \hbar^{4}}{2m^{2}}[\int d^{2}r
\rho (r) (y \frac{\partial}{\partial x} + x \frac{\partial}{\partial y})^{2}
\rho (r) \\ \nonumber & + & \int d^{2}r (x \frac{\partial \rho}{\partial y} +
y \frac{\partial \rho}{\partial x})^{2}].
\end{eqnarray}
 
Using the variational wave function, we can get all these moments. In this
case $ m_{3}(ee) $ exactly vanishes. So the frequency for the quadrupole mode
is,
\begin{equation}\label{qu2} 
\frac{\omega_{s}^2}{\omega_{0}^2} = \frac{4
\lambda \eta^2 }{(2 + P \eta )} + (\lambda + \frac{1}{\lambda }). 
\end{equation} 
This expression for the quadrupole frequency is also the same as
Eq.(\ref{qu1}). So $ \omega_{-} $ in Eq. (\ref{mono2}) and $ \omega_s $ in
Eq.(\ref{qu2}) shows the splitting occurs between two
quadrupole modes in a two dimensional deformed trapped Bose gas and the
dependence of the splitting on the interaction strength and trap anisotropy
can be analyzed from the analytical expressions. For an isotropic trap, the
two quadrupole modes are degenerate. 
The variation of the splitting between two quadrupole modes $ \Delta_{b}=
\omega_s - \omega_{-} $, of a trapped interacting Bose gas, with the
dimensionless interaction parameter $P$ is shown in Fig. 2. 

We have checked that the sum rule method gives correct results for the
excitation frequencies of the two quadrupole modes for a system of
interacting bosons in an anisotropic trap. Now we apply this method to
calculate the excitation energies of the quadrupole modes of a system of
interacting Fermions in a deformed trap. 
             
\section{two dimensional trapped anisotropic Fermi gas at zero temperature}

In this section we discuss the collective oscillation of a two dimensional
deformed trapped unpolarized Fermi gas at zero temperature within the
sum-rule approach.  Using this approach, the collective excitations have been
studied in other finite Fermionic systems like atomic nuclei, metal clusters
\cite{brack} and quantum dots \cite{sinha}. 

We consider a two dimensional deformed trapped unpolarized Fermionic atoms at
very low temperature. The two-body interaction of the dilute gas can be
described by the short range pseudopotential $ V(\vec r - \vec r^{\prime}) =
g_2 \delta^2 (\vec r - \vec r^{\prime})$, where $ g_{2} $ is the coupling
constant and its form is given in Sec.II. The Hamiltonian of the trapped
Fermionic atoms is given by, 
\begin{equation}\label{ham} 
H = \sum_{i} \frac{
p_{i}^2 }{2 m} + V_{ext} + g_2 \sum_{i < j}
 \delta^{2} ( \vec r_i - \vec r_j ) ,
 \end{equation}
 where the confining potential is 
 \begin{equation}
 V_{ext} = \frac{1}{2} m \omega_{0}^2 (\lambda x^2 + \frac{ y^2}{\lambda} ).
\end{equation}

The Thomas-Fermi energy functional of this trapped interacting Fermi system
is given by, 
\begin{equation} 
E[\rho (r)] = \int d^2 r [\frac{\hbar^2 \pi
}{2m} \rho^2(r) +
 V_{ext} \rho(r) + \frac{ \tilde{g_2}}{2} \rho^2(r)],
\end{equation} where $
\tilde{g_{2}} = g_{2}/2 $. Here we assume the density of two spin components
are same, $\rho_{\uparrow} = \rho_{\downarrow}$. The interaction energy
density $g_{2} \rho_{\uparrow} \rho_{\downarrow}$, can be written as
$\frac{g_{2}}{4}\rho^{2}$, where $\rho$ is the total density. By minimizing
the energy functional with respect to density, we obtain,
\begin{equation}\label{den} 
\rho(\vec r) = \frac{ R_{F}^2 }{2 K_0 \pi a_{0}^4
} [ 1 - \frac{ r^2} {R_{F}^2}] , r \leq R_{F}, 
\end{equation} 
where $ R_{F} =
(4N K_0 )^{1/4} a_0 $ is the radius of the atomic gas which is determined
from the condition $ \int d^2r \rho (r) = N $.  $ K_0 = 1 + \frac{\tilde{g_2}
m }{\pi \hbar^2} $ is a dimensionless constant. At very low temperatures,
collisions are suppressed due to Fermi statistics and system is in the
collisionless regime. We study the collective excitation frequencies in this
regime by sum rule approach. 

In Sec.III we have derived the expressions for quadrupole excitation
frequencies within sum-rule approach. We can use  expression
Eq.(\ref{sum1}), to calculate one of the quadrupole mode frequencies for Fermi
gas also. 

We evaluate all the expectation values of the corresponding operators by
using the Thomas-Fermi density (\ref{den}),
 \begin{equation}\label{both}
\frac{E_{x}}{<x^2>} \pm \frac{E_{y}}{<y^2>} = \frac{m \omega_{0}^2}{2}
 \frac{ (3 K_0 + 1)}{2 K_{0}}(\lambda \pm \frac{1}{\lambda}). 
\end{equation}
Using Eqs. (\ref{sum1}) and (\ref{both}), we obtain,
\begin{eqnarray}\label{mono3} 
\frac{\omega_{\pm}^2}{\omega_{0}^2} & = &
\frac{(3 K_0 + 1)}{2 K_{0}}(\lambda + \frac{1}{\lambda})
 \\ \nonumber & \pm & \sqrt{[ \frac{(3 K_0 + 1)}{2 K_{0}} (\lambda -
\frac{1}{\lambda })]^2 + (1 - \frac{ 1 }{ K_0 })^2}.
 \end{eqnarray} 
For an
isotropic trap, the monopole mode frequency becomes $ \omega_{+} = 2
\omega_{0} $. 

There is another quadrupole mode for which the excitation operator is $ F = x
y $. Using the density for the trapped interacting Fermi gas at $ T = 0 $, we
get the following moments: 
\begin{equation} 
m_{1} = \frac{ \hbar^3 R_{F}^6
}{48 m^2 a_{0}^6 \omega_{0} K_0 } (\lambda + \frac{1}{ \lambda }),
\end{equation} 
\begin{equation} 
m_{3} = \frac{ \hbar^5 \omega_{0} R_{F}^6
}{12 K_0 m^2 a_{0}^6 } [ \frac{1}{K_0} + \frac{(\lambda + \frac{1}{ \lambda
})^2}{4}].
\end{equation} 
In this case also, $ m_3(ee) $ exactly vanishes. The
quadrupole oscillation frequency is given by, 
\begin{equation}\label{qu3}
\frac{\omega_s}{\omega_0} = \sqrt{ [(\lambda + \frac{1}{\lambda }) +
\frac{4}{K_0 (\lambda + \frac{1}{\lambda })}]}.
 \end{equation} 
For an
isotropic trap, $ \omega_{-} $ in Eq. (\ref{mono3}) and Eq. (\ref{qu3})
becomes 
\begin{equation} 
\omega_{s} = \sqrt{2} \omega_{0} \sqrt{\frac{ (1 +
K_0)}{K_0}}. 
\end{equation} 
In Eq.(\ref{mono3}) and Eq.({\ref{qu3}), $
\omega_{-} $ exhibits the splitting of the quadrupole modes of a two
dimensional deformed trapped Fermi gas. The scissors mode is also discussed
in Ref. \cite{tosi} for superfluid Fermi gas. 

The monopole mode frequency of an isotropic trapped interacting Fermi system
is $ 2\omega_0 $ which is independent of the interaction strength of the
two-body potential and the number of particles. This is because of the
presence of SO(2,1) symmetry in the Hamiltonian (Eq.(\ref{ham})) as discussed
in the previous section. 

The splitting between two quadrupole modes, $\Delta_{f} = \omega_{s} -
\omega_{-}$ , of a deformed trapped interacting Fermi gas is shown in Fig. 3.
The frequencies of these two modes and the splitting between them are
independent of the particle number for two dimensional Fermions within 
Thomas-Fermi approximation. This splitting decreases almost linearly with 
increasing interaction strength. 

\section{summary and conclusions}

In this paper, we have mainly considered  two non-degenerate quadrupole modes
of a quantum gas in an anisotropic harmonic oscillator potential. We
investigated the effect of interaction on the splitting between these
quadrupole modes for arbitrary trap deformation. We have calculated a few
low-lying collective excitation frequencies of a two dimensional trapped Bose
gas in an anisotropic trap, by using time dependent variational method. We
found that one quadrupole mode is coupled with the monopole mode in presence of trap 
deformation. Another quadrupole mode associated  with the fluctuation of the average value 
of $xy$ (which is also known as scissors mode), is decoupled. 

Using the energy weighted sum-rule approach we derived the general dispersion
relation of the two quadrupole excitations. Using the same variational wave
function for Bosons, we checked that the collective frequencies obtained from
the sum-rule approach are exactly the same as those obtained from the
variational method. The main advantage of the sum-rule method is that it can
be applied to both trapped Bosons and Fermions to calculate the excitation
frequencies in the collisionless regime. This method can be applied for any
number of confined particles and also it can be generalised for long range
interactions. This energy weighted sum-rule method can be extended  for Coulomb
interaction to study the quadrupole excitations in a deformed electronic
nanostructure like an elliptic quantum dot.  The splitting between the
quadrupole modes obtained from this method is non perturbative in the trap
anisotropy parameter. 

We considered a system of two dimensional spin unpolarised interacting
Fermions in an anisotropic harmonic oscillator potential within Thomas-Fermi
approximation. Applying the sum-rule technique to this deformed Fermi gas, we
obtain two quadrupole excitation frequencies and the splitting between them
analytically. For both statistics, the amount of splitting between the
quadrupole modes decreases with  increasing interaction strength. For a two
dimensional Fermi system the frequencies and the splitting are independent of
the particle number. For an isotropic trap, the monopole mode frequency of a
Bose gas as well as Fermi gas is the universal frequency $ 2 \omega_0 $. This
monopole mode frequency is independent of the strength of the two-body
interaction potential and the number of particles. This is due to the underlying $ 
SO(2,1) $ symmetry in the Hamiltonian. Strictly speaking, our
all the results are valid when the conditions $ \hbar \omega_z \gg \mu \geq \hbar
\omega_0 $ and $ k_B T \ll \hbar \omega_z $ are satisfied.

Recent experimental progress in MIT \cite{gorlitz} on
quasi two dimensional Bose condensed shows the possibilities of verification of
our results.
Above mentioned quadrupole modes are excited in the two dimensional plane and 
for simplicity we consider only the two dimensional trapped gas. This method and the most
general Gaussian anstaz for the order parameter can also be extended to three 
dimensional anisotropic systems to study the various quadrupole modes.
The splitting in these two quadrupole modes may be used to find trap
anisotropy. It will be an interesting to study the splitting between the quadrupole modes 
of an anisotropic quantum system in presence of terms having definite chirality, like 
magnetic field or rotation. 

We would like to thank G. Baskaran and M. V. N.
Murthy for helpful discussions. LKB is a unit\'e de recherche de l'Ecole
normale sup\'erieure et de l'Universit\ 'e Pierre et Marie Curie, associ\'ee
au CNRS.


\begin{figure}

\caption{ The variation of the ratio of the widths of the 
condensate, $ \eta$ as a function of the dimensionless
effective interaction strength $ P $, for the fixed ratio of trap frequencies 
$ \lambda = 0.7 $.}

\caption{ The difference between the two 
quadrupole modes of an interacting Bose gas, $ \Delta_{b}/
\omega_{0} $ as a function of the 
dimensionless effective interaction strength $ P $ for fixed ratio of trap frequencies
$ \lambda = 0.7 $.}

\caption{ The difference between the two 
quadrupole modes of an interacting unpolarized Fermi gas $
\Delta_{f} / \omega_{0} $, as a function 
of the dimensionless parameter $ K_{0} $ for fixed ratio of trap frequencies 
$ \lambda = 0.7 $.}

\end{figure}


\begin{thebibliography}{99}

\bibitem{anderson} 
M. H. Anderson, J. R. Ensher, M. R. Matthews, C. E.
Wieman, and E. A. Cornell, Science {\bf 269}, 198 (1995); C. C. Bradley, C.
A. Sackett, J. J. Tollet, and R. G. Hulet, Phys. Rev. Lett. {\bf 75}, 1687
(1995); K. B. Davis, M. O. Mewes, M. R. Andrews, N. J. van Druten, D. S.
Durfeee, D. M. Kurn, and W. Ketterle, Phys. Rev. Lett. {\bf 75} 3969 (1995). 

\bibitem{dal} 
F. Dalfovo, S. Giorgini, L. P. Pitaevskii, and S. Stringari,
Reviews of Modern Physics, {\bf 71}, 463 (1999); A. S. Parkins and D. F.
Walls, Phys. Rep. {\bf 303}, 1 (1998); A. Griffin, Preprint, cond-mat/
9911419. 
 
\bibitem{strin} 
S. Stringari, Phy. Rev. Lett. , {\bf 77}, 2360 (1996). 

\bibitem{gar}
 Victor M. Perez-Garcia, H. Michinel, J. I. Cirac, M. Lewenstein, and P.
Zoller, Phys. Rev. Lett. {\bf 77}, 5320 (1996). 

\bibitem{castin} Y. Castin and R. Dum, Phys. Rev. Lett. {\bf 77}, 5315
(1996). 

\bibitem{kagan} Y. Kagan, E. L. Surokov and G. Shlyapnikov, Phys. Rev. A.
{\bf 54 }, R1753 (1996). 

\bibitem{olsha}
 Pippa Storey and M. Olshanii, Phys. Rev. A. {\bf 62}, 033604 (2000). 

\bibitem{mew}
D. S. Jin, J. R. Ensher, M. R. Matthews, C. E. Wiemann, and E. A. Cornell,
Phys. Rev. Lett. {\bf 77 }, 420 (1996);

 M. O. Mewes, M. R. Andrews, N. J. van Druten, D. M. Kurn, D. S.
Drfee, C. G. Townsend, and W. Ketterle. Phys. rev. Lett. {\bf 77}, 988
(1996). 
 
\bibitem{jin}
D. S. Jin, M. R. Matthews, J. R, Ensher, C. E. Wiemann, and E. A. Cornell,
Phys. Rev. Lett. {\bf 78 }, 764 (1997).

\bibitem{jin1} 
B. DeMacro and D. S. Jin, Science {\bf 285}, 1703 (1999);  M.
O. Mewes, G. Ferrari, F. Schreck, A. Sinatra and C. Salomon, Phys. Rev. A
{\bf 61} ,011403(R) (2000). 
 
\bibitem{butt}
 D. A. Butts and D. S. Rokhsar. Phys. Rev. A {\bf 55}, 4346
(1997). 

\bibitem{salas}
 L. Salasnich, B. Pozz, A. Parola and L. Reatto, J. Phys. B
{\bf 33 }, 3943 (2000). 

\bibitem{amo} M. Amoruso, I. Meccoli, A. Minguzzi and M. P. Tosi, The Euro.
Phys. Jour. D, {\bf 7}, 441 (1999). 


\bibitem{vichi}
L. Vichi and S. Stringari.Phys. Rev. A {\bf 60}, 4734 (1999).

\bibitem{clark}
Georg M. Brunn and Charles W. Clark, Phys. Rev. Lett. {\bf 83}, 5415 (2000).


\bibitem{tosi}
A. Minguzzi and M. P. Tosi, cond-mat/0005098.

\bibitem{baranov}
M. A. Baranov and D. S. Petrov, Phys. Rev. A, {\bf 62}, 041601 (2000).

\bibitem{petrov}
 T. Haugset and H. Haugerud, Phys. Rev. A. {\bf 57}, 3809 (1998).
 Sadhan K. Adhikari, Phys. Rev. E {\bf 62}, 2937 (2000); 
 Tarun K. Ghosh. Phys. Rev. A, {\bf 63}, 013603 (2001);.

\bibitem{ma}
Tin-Lun Ho and Michael Ma, J. Low Temp. Phys. {\bf 115}, 61 (1999).

\bibitem{pit}
L. P. Pitaevskii and A. Rosch, Phys. Rev. A, {\bf 55}, R853, (1997);
Pijush K. Ghosh, Phys. Rev. A, {\bf 65}, 012103 (2002).

\bibitem{kanti}
Tarun K. Ghosh, Phys. Lett. A, {\bf 285}, 222 (2001). 

\bibitem{piju}
Pijush K. Ghosh, cond-mat/0109073.

\bibitem{sank}
Tin-Lin Ho, C. V. Ciobanu. Phys. Rev. Lett. {\bf 85 }, 4648 (2000);
Sankalpa Ghosh, M. V. N. Murthy and Subhasis Sinha. Phys. Rev. A,
{\bf 64}, 053603 (2001).

\bibitem{gorlitz}
A. Gorlitz {\em et al}, Phys. Rev. Lett. {\bf 87}, 130402 (2001).

\bibitem{burger}
S. Burger {\em et al}, cond-mat/0108037.

\bibitem{gross}
E. P. Gross, Nuovo Cimento A {\bf 20}, 454 (1961); L. P. Pitaevskii, Pis'ma Zh. Eksp.
Teor. Fix. {\bf 77}, 988 (1961) [ Sov. Phys. JETP {\bf 13 },451 (1961)].

\bibitem{baganato}
V. Baganato and D. Kleppner, Phys. Rev. A, {\bf 44}, 7439  (1991).


\bibitem{guery}
D. Guery-Odelin and S. Stringari, Phys. Rev. Lett. {\bf 83}, 4452 (1999).

\bibitem{mara}
O. M. Marago, S. A. Hopkins, J. Arlt, E. Hodby, G. Hechenblaikner and
C. J. Foot, Phys. Rev. Lett. {\bf 84 }, 2056 (2000).

\bibitem{bohi}
O. Bohigas, A. M. Lane, J. Martorell, Phys. Rep. {\bf 51}, 267 (1979);
E. Lipparini and S. Stringari, Phys. Rep. {\bf 175 }, 102 (1989).

\bibitem{lip}
E. Lipparini, N. Barberan, M. Barranco, M. Pi, L.Serra. Phys. Rev. B, {\bf 56}, 12375 (1997).

\bibitem{brack}
W. de Heer, Rev. Mod. Phys. {\bf 65 }, 611 (1993);
M. Brack, Rev. Mod. Phys. {\bf 65 }, 677 (1993).

\bibitem{sinha}
Subhasis Sinha, Physica E, {\bf 8}, 24 (2000).


\end{thebibliography}
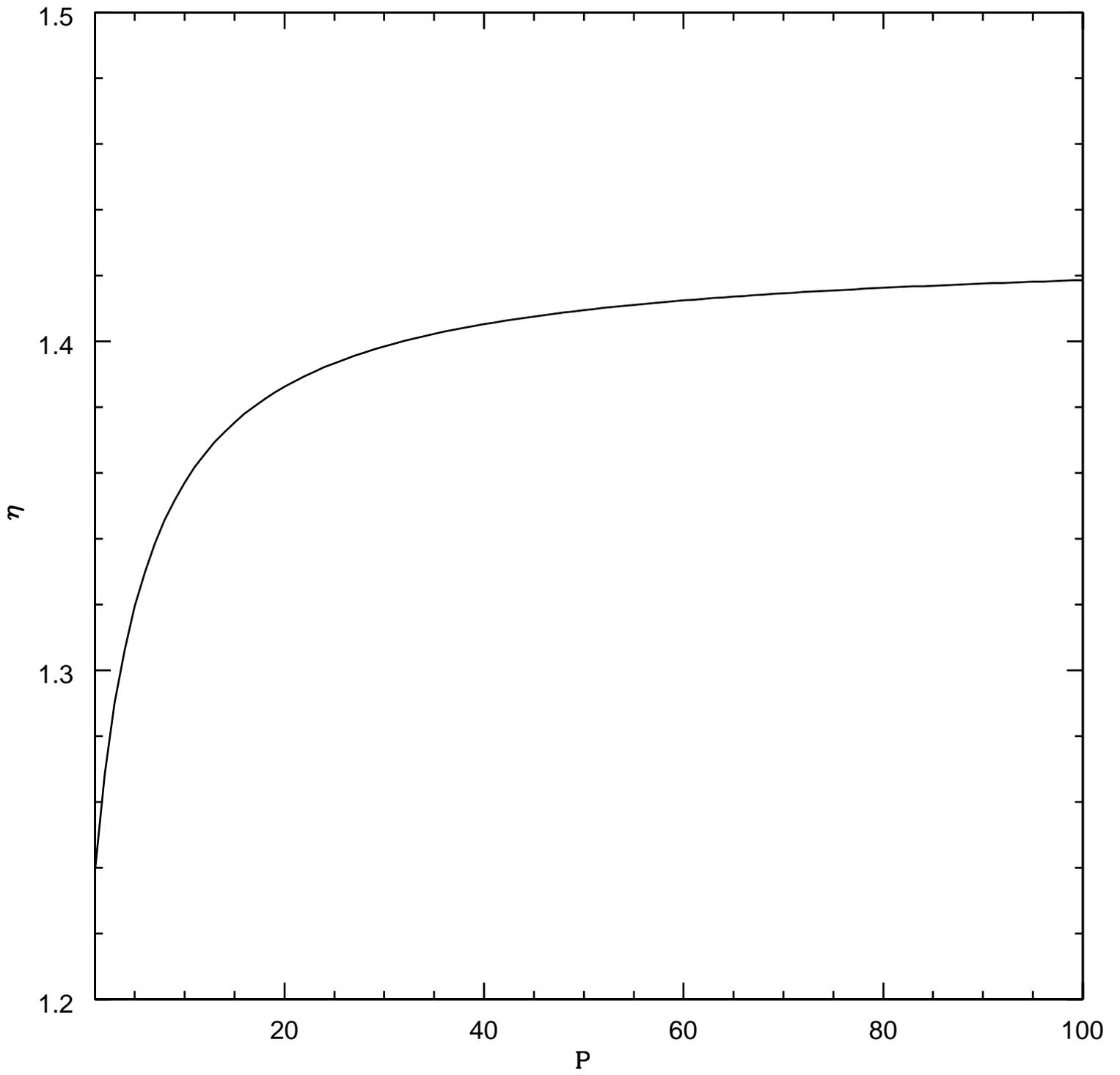
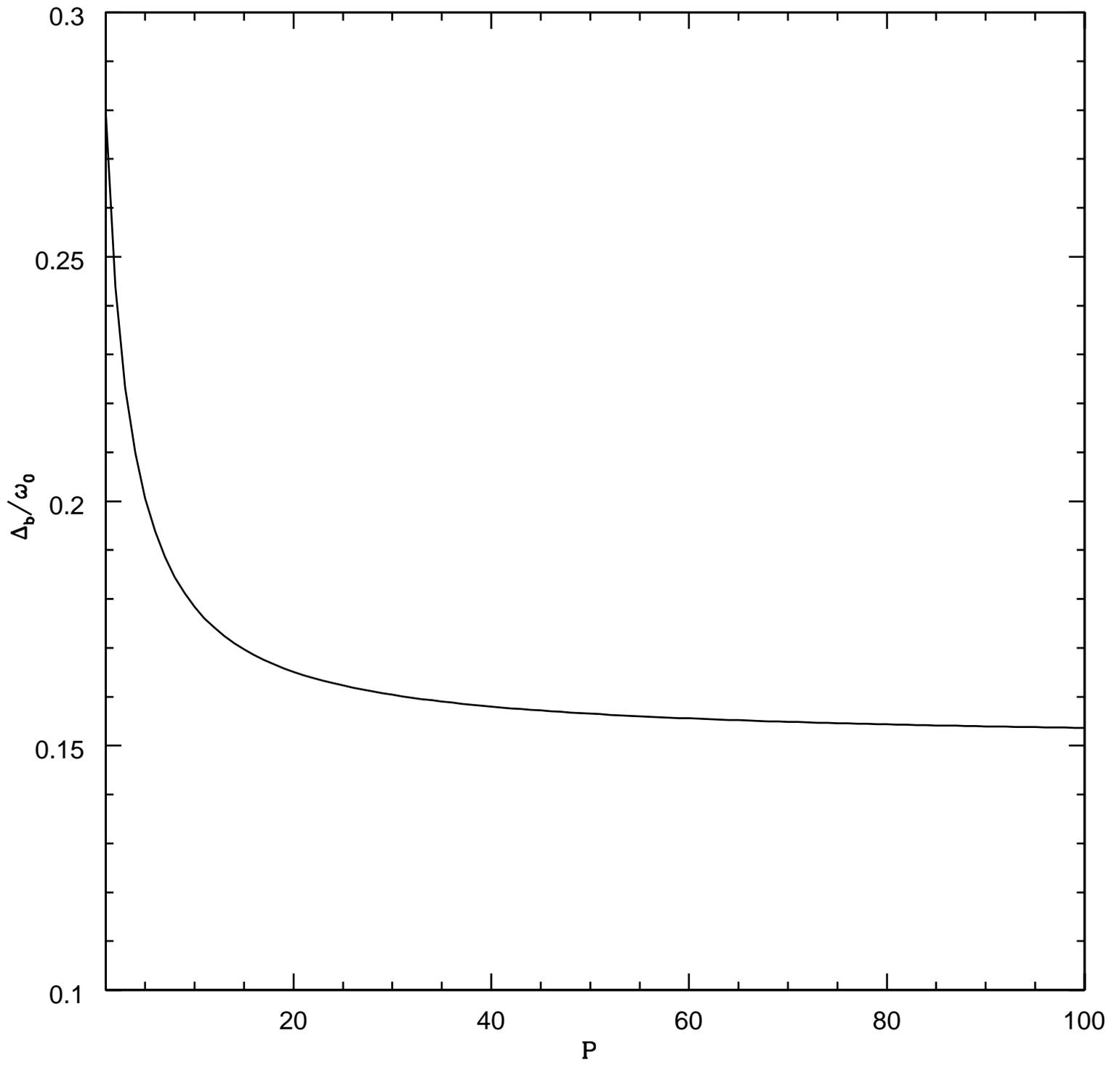
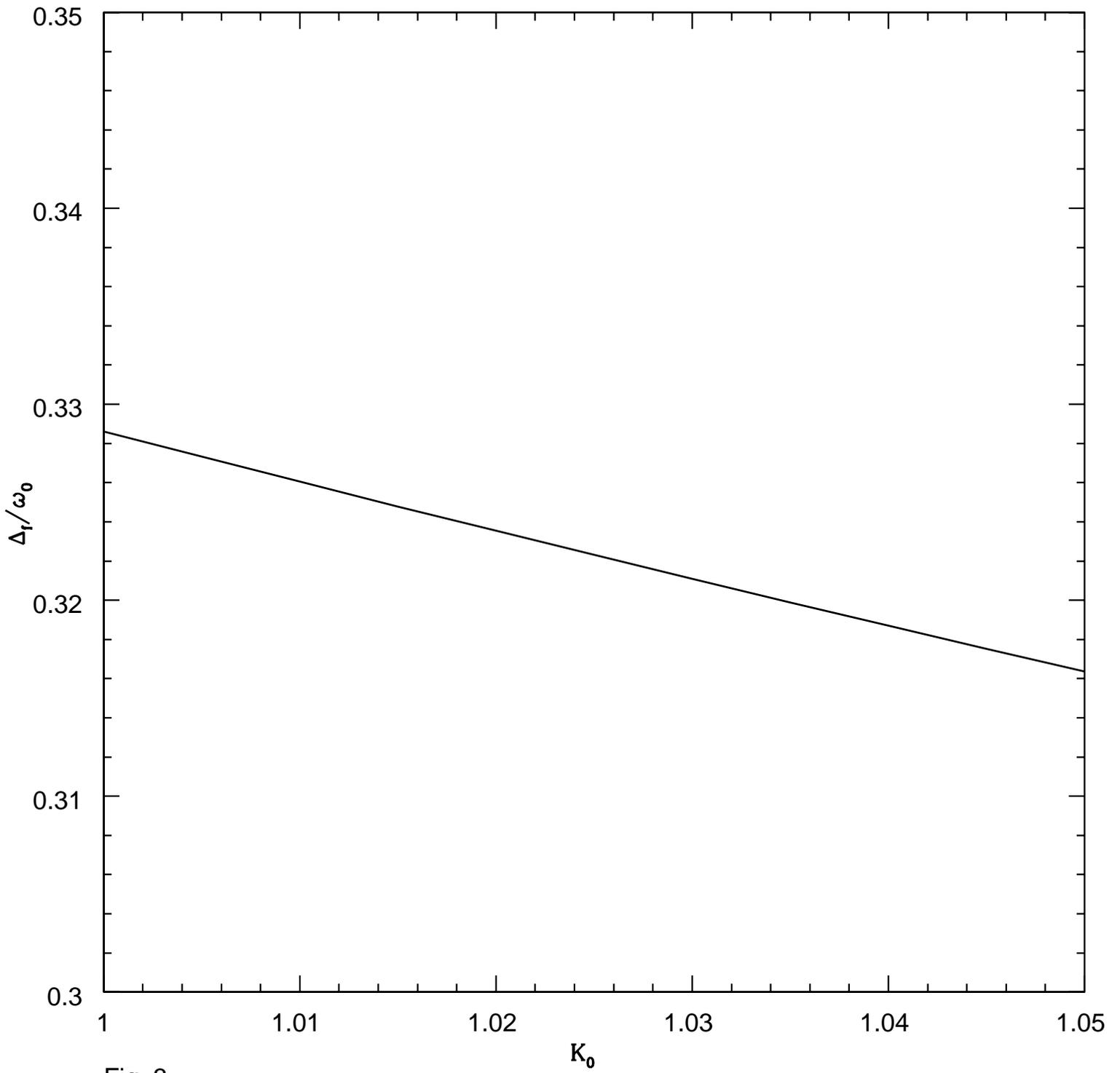
\end{document}